\documentclass[12pt]{article}

\usepackage{amsmath}
\usepackage{amsfonts}

\begin{document}
\title{\huge{Current Trends in Mathematical Cosmology}}
\author{
{\Large Spiros Cotsakis}\\
GEODYSYC\\
Department of Mathematics\\
University of the Aegean\\
Karlovassi, 83 200, Greece\\
$\mathtt{email: skot@aegean.gr}$
}
\maketitle
\begin{abstract}
\noindent We present an elementary account of mathematical cosmology through
a series of important unsolved problems. We introduce the
fundamental notion of \emph{a cosmology} and focus on the issue of
singularities as a theme  unifying  many current, seemingly
unrelated trends of this subject. We discuss problems associated
with the definition and asymptotic structure of the notion of
cosmological solution and also problems related to the
qualification of approximations and to the ranges of validity of
given cosmologies.
\end{abstract}

\section{Introduction}
\begin{quote}\emph{Circumstancial evidence is a very
tricky thing. It may seem to point very straight to one thing, but
if you shift your own point of view a little, you may find it
pointing in an equally uncompromising manner to something entirely
different. (Sherlock Holmes)}\end{quote} In any field of applied
mathematics one starts by carefully identifying the basic object of
study, one that contains the essential parameters of the problem,
to be determined by later analysis. The mathematical methods of
cosmology which promise to be useful, even essential to the
nature of problems one typically encounters come from two
different sources, namely, differential geometry and dynamical
systems theory. Indeed, \emph{mathematical cosmology}, a largely
unexplored but highly interesting and promising area of research,
may be loosely defined as that separate discipline in applied
mathematics which lies in the differentiable world of geometry and
dynamical systems borrowing heavily from both areas and
contributing back constantly new problems and ideas not only to
both of these mathematical fields but also to the closely related
physical and observational cosmology.

The basic object of study in any mathematical approach to
cosmological problems is that of \emph{a cosmological model}. Let
us explain briefly what a cosmological model is and how we can
generate interesting models. 
We call \emph{a
cosmology} the result of combining the mathematical theorizing
that goes into the construction of a cosmological model with the
observational data that are available in the astronomical
literature. In the following, however, we shall ignore this
difference between a cosmology and a (cosmological) model and use
both words indistinguishably to describe this fundamental notion
of cosmological modelling.

In this paper we lay the foundations of mathematical cosmology in
a manner suitable for the nonspecialist, focusing on the
fundamental mathematical problems which single out this field as a
separate component within applied mathematics and mathematical
physics. The presentation is elementary and is addressed to those
who need a general overview before plugging in the excruciating
details.

In the next Section, we introduce the idea of a cosmology as a
basic unknown of this subject. Section 3 discusses the notion of
cosmological law and shows how the singularity problem, a central
issue in this field, is used to orient the whole of mathematical
cosmology research around three basic themes, namely, global
evolution, approximations and range of validity of a cosmology.
Sections 4 to 6 describe in more detail these basic avenues of
research expanding on several open questions relevant to each
theme. We conclude in Section 7 with some more general  comments
on the nature of mathematical modelling in cosmology.

This is meant to be a short review of a huge subject and therefore
we apologize in advance for many superficial passages, or possible
omissions of important ideas and works by fellow
mathematical cosmologists over the past decades. In this sense the
bibliography contains some works which the author has found
relevant in the preparation of this paper and is only meant to be
a useful guide to those interested in pursuing this beautiful
subject further. It contains mainly review articles and books and
is not to be regarded as a declaration of the most important
sources in our field.

\section{Cosmologies}
\begin{quote}\emph{There is nothing so unnatural as the commonplace.
(Sherlock Holmes)}\end{quote} There are three essential elements
that go into a cosmology:
\begin{itemize}
\item  A cosmological spacetime $\mathtt{(CS)}$
\item  A theory of gravity $\mathtt{(TG)}$
\item  A collection of matterfields $\mathtt{(MF)}$
\end{itemize}
A cosmology is a particular way of combining these three basic
elements into a meaningful whole:
\begin{equation}
\mathtt{Cosmology = CS+TG+MF}. \label{1}%
\end{equation}
There is a basic hierarchy of $\mathtt{CS}$s according to the
degree of exact symmetry present. We basically start with a smooth
manifold $\mathcal{M}$ and impose a Lorentzian metric $g_{ab}$ on
$\mathcal{M}$ which admits a number of symmetries. Generally
speaking, the $\mathtt{CS}$'s hierarchy list is:
\begin{enumerate}
\item  Isotropic (Friedmann-Robertson-Walker) spacetimes
\item  Homogeneous (Bianchi) spacetimes
\item  Inhomogeneous spacetimes
\item  Generic spacetimes
\end{enumerate}
This list is one of decreasing symmetry, and so increasing
generality, as we move from top to bottom and comprises four
families of $\mathtt{CS}$s. The last family, generic spacetimes,
has no symmetry whereas the isotropic spaces correspond to the
simplest (and perhaps unphysical), highest-symmetry toy models
that exist.

$\mathtt{TG}$ too, fortunately or not depending on how one looks
at it, come in great variety. A partial list of important
families of theories which include gravity-a necessary ingredient
for the modern construction of cosmologies-is:
\begin{enumerate}
\item  General relativity (GR)
\item  Higher derivative gravity theories (HDG)
\item  Scalar-tensor theories (ST)
\item  Superstring theories (SS)
\end{enumerate}
It is widely accepted today, after the pioneering work of Hawking,
Geroch and Penrose in the late sixties (see \cite{ha-el73}
 for an account of these results) that GR leads to
singularities in the early development of generic $\mathtt{CS}$s
and consequently one needs a better $\mathtt{TG}$ to account for
early cosmological events in a consistent way. The above list is
motivated partially by these results and comprises, besides GR,
modifications involving higher derivatives, scalarfield-curvature
couplings as well as supersymmetric ideas in the formulation of
entries 2, 3 and 4 above respectively.

Matterfields also come in an ambitious shopping list of
interesting candidates which may have played an important role
during different epochs in the history of the universe. For
example, we can consider:
\begin{enumerate}
\item  Vacuum
\item  Fluids
\item  Scalar fields
\item $n-$form fields
\end{enumerate}
See Table \ref{table1} for a summary. Definition (\ref{1}) above
is a very broad one. The simplest and best studied (relativistic)
cosmology of physical interest is the (FRW/GR/Fluid) cosmology.
\begin{table}[ht]
\caption{Three essential elements comprising a
cosmology according to Eq. (\ref{1}). Also shown are several
members of each particular element.}
\begin{center}
\begin{tabular}{l|c|r}
\multicolumn{3}{c}{\emph{\textbf{Cosmologies}}}\cr\hline {\bf
Theories of gravity} & \textbf{Cosmological spacetimes} &
\textbf{Matterfields}\cr\hline\hline General Relativity &
Isotropic & Vacuum\cr Higher Derivative Gravity & Homogeneous &
Fluids \cr Scalar-Tensor theories & Inhomogeneous & Scalarfields
\cr String theories & Generic & $n-$form fields \cr\hline
\end{tabular}
\end{center}\label{table1}
\end{table}This is in fact the cosmology discussed in many textbooks on the
subject under the heading `Relativistic Cosmology',  but we may
obviously attempt to construct and analyze other possible
cosmologies. For example, we can consider the families:
\begin{itemize}
\item  FRW/GR/vacuum
\item  Bianchi/ST/fluid
\item  Inhomogeneous/String/$n$-form
\item  Generic/GR/vacuum
\end{itemize}
\noindent and so on. For visualization purposes, we can consider
the 3-dimensional 'space' of all cosmologies with coordinates
($\mathtt{CS}$,$\mathtt{TG}$,$\mathtt{MF}$) the `points' of which
represent different cosmologies\footnote{We are well aware of the
danger present here of being too pedantic or intimidating for most
readers when discussing such `meta-cosmological' issues. However,
if one wishes to ponder for a minute about the difficulties
involved in the construction of such a space, we note that, to the
best of our knowledge, it is unclear at present how to put an
ordering in the three axes $\mathtt{CS}$, $\mathtt{TG}$ and
$\mathtt{MF}$. The older notion of superspace (cf.  \cite{wh68})
 provides such an order relation only for the
$\mathtt{CS}$-axis of our cosmology-space. The construction of
analogous orderings for lagrangians corresponding to $\mathtt{TG}$
and $\mathtt{MF}$ leading to a consistent topology on the space of
all cosmologies is beyond the scope of the present article and
could constitute an interesting avenue of research.}. For example,
the category ($\cdot$/GR/$\cdot$), a higher dimensional subspace in the basic cosmology
space, is the best studied cosmology so far -- see the excellent
review \cite{el-el98}.

\section{Global evolution, approximations and range of validity}
\begin{quote}
\emph{Functions, just like living beings, are characterized by
their singularities. (P. Montel)}
\end{quote}
The tool we use to translate the above into a consistent
mathematical language is the Action Principle. We use this tool to
formulate precisely the notions of a $\mathtt{TG}$ and that of a
$\mathtt{MF}$. So, how do we construct a cosmology? Pick up a
spacetime from the cosmological hierarchy list, choose a gravity
theory and one or more matterfields, tie them together through the
Action Principle and try to explain the observed facts in terms of
the consequences of the application of the variational principle
(for a detailed mathematical introduction see \cite{co01}).

Through the action principle, the resulting \emph{cosmological
equations} one obtains by starting from a \emph{cosmological
lagrangian} and using symmetry or other phenomenological
considerations contain the basic properties, to be unraveled, of
any cosmology. There are many questions that can be asked for any
such set of equations, leading in this way to many different
fundamental trends in theoretical cosmology today and of course to
the rest of this paper. Before we proceed to discuss  some of
these problems, however, we pause to explain what a cosmological
equation has to do with another basic notion that we shall
encounter, \emph{the cosmological law}.

We distinguish, for the purpose of orientation, two kinds of
such laws, that is, \emph{fundamental} and \emph{effective}
cosmological laws\footnote{There is another kind of cosmological
`law', the set of ideas that goes by the name of \emph{The
Anthropic Principle} \cite{ba-ti86}. However, the
usage of the word `law' we adopt here only includes those that are
formulated in the form of dynamical systems. It is, we believe, an
intriguing question whether the Anthropic Principle has some
hidden dynamical meaning and, if yes, how could this be possibly
framed in the form of a differential equation.}. We talk of a
fundamental cosmological law when we are faced with a set of
equations of the form ($\mathtt{TG}$/$\mathtt{MF}$) that is, when
we have not imposed any symmetry in the underline spacetime. Now
it is probably somewhat surprising or even misleading to call,
say, the full Einstein equations, $G_{ab}=kT_{ab}$, a
\emph{cosmological} law for, any such set of equations contains
much more than cosmological solutions eg., it contains black holes
or gravitational waves. The only justification for this
terminology is that in the full, `unconstrained', case one is
interested in the behaviour of the \emph{whole} spacetime and this
does not have \emph{a priori} imposed on it any specific
assumption that would lead to other kinds of solutions (for
example asymptotic flatness etc). Hence, a
fundamental cosmological law has only mild assumptions on $(\mathcal{M}%
,g_{ab})$, for instance it can be taken to be globally hyperbolic with only
some physically reasonable energy condition imposed on the matter content.

On the other hand any imposition of symmetry or other reduction
principle on $(\mathcal{M},g_{ab})$ and the matter fields leads to
\emph{effective cosmological laws}, that is, to more special sets
of differential equations which are thus obtained as byproducts of
a given fundamental cosmological law. Thus for instance, imposing
on the fundamental law (GR/matter) the usual isotropicity
conditions and requiring the admission of a perfect fluid matter
source we obtain the effective law (FRW/GR/Perfect Fluid) -- a set of
equations for the time evolution of the scale factor and the fluid
parameters.

Even at this stage we easily realize the importance of genericity
vs. symmetry or fundamental vs. effective cosmological laws. This
dichotomy raises a basic question of principle: Taking for granted
the extreme difficulty to handle mathematically any fundamental
law to produce strong results, how could we ever be sure that we
obtained reliable or generic results while working at the `lower'
levels of effective laws? Soon we shall be more specific and
have more to say.

We now return to our basic theme. The problem of possible
break-down or singularity in the future or past of any given
cosmology is the most basic problem in mathematical cosmology and
affects all cosmologies. (Almost) all cosmological solutions are
likely to form singularities in a finite time.  This forces us to
consider the following two fundamental questions which frame the
singularity problem in cosmology: `What do we mean by a
cosmological solution?' and, `What is the range of validity of a
given cosmology?'. The singularity problem  has several
interelated offshoots:
\begin{itemize}
\item  Where are the cosmological singularities to be found?
\item  Why do cosmological solutions have the tendency to develop singularities?
\item  What is the nature of the cosmological singularity?
\item  Can we continue the solution past the singularity?
\end{itemize}
Indeed the singularity problem can be efficiently used to
signpost the current status of our subject. What is then the
present state of mathematical cosmology? Overall our present
efforts are directed to
\begin{enumerate}
\item \textbf{Global evolution:} \emph{Understand the global evolution of
solutions to the cosmological equations resulting from all possible available laws}
\item \textbf{Qualifications of approximations:} \emph{Evaluate the various
approximations (especially matterfields) involved}
\item \textbf{Range of validity:} \emph{Decide on the range of validity of the
various cosmologies and clarify the meaning of the notion of cosmological solution}
\end{enumerate}
In the next three Sections we take up in turn each one of the
above fundamental trends and present in some detail some of the
basic sub-topics that the community of mathematical cosmologists
has found interesting and occupied itself with over the years.
\section{Asymptotic cosmological states}
\begin{quote}\emph{  If there is no time, there is no
space. (Yvonne Choquet-Bruhat)} \end{quote} The issue of
determining the global cosmological evolution, \emph{alias} the
problem of the asymptotic cosmological states, is a very basic and
by and large open research problem in mathematical cosmology
today. In view of the impossibility of meaningfully framing
universal boundary conditions in cosmology this problem becomes
particularly important in any attempt to understand the long term
behaviour of cosmological systems. Its two components, dynamics in
the positive, expanding direction and that in the negative or
contracting direction present us with different issues,  not least
 because of their different physical interpretation.

The dimensionality of cosmological dynamical systems varies from
one (in the case of the simplest (FRW/GR) cosmologies) to infinity
(generic cosmologies) and most of them are typically formulated in
higher than two dimensions. This, together with the essential
nonlinearities present in any cosmological law from which these
systems are derived, results in making an already difficult
subject even more demanding (and interesting!). Attempting to
reconstruct and classify the known results in the dynamics of
cosmology according to dimension, gathering all systems of equal
dimensionality together and discovering common features (`what do
we know about 1D cosmologies, 2D cosmologies, etc?') might be an
interesting project.

A general feature of the dynamics of cosmologies in the
contracting direction is that things typically tend to  become
more complicated. The well-known BKL approximation scheme for approaching the
singularity is the typical example \cite{bkl70,bkl82}. Qualifying the dynamics of
contracting cosmologies has been a central problem in mathematical
cosmology for many years. The pioneering work of Barrow in the
early eighties on (Bianchi/GR) contracting cosmologies  \cite{ba82}
 established the connection between the complicated
patterns of oscillations present in BKL and in the hamiltonian
picture of Misner \cite{mi94} and the theory of chaotic dynamical
systems,  thus opening up a whole new chapter in mathematical
cosmology. After these works there has been a large body of
literature connected with the issue of chaotic behaviour in
different cosmologies continuing even to this day with many open
problems still remaining (see also the following Section).

The behaviour of cosmologies in the expanding direction, however,
appears to be of a completely different nature at least as far as
the types of questions with which one is concerned. Chaotic behaviour in
the future does not appear to be a typical feature of an expanding
cosmology. Instead one is content in asking questions having to
do with \emph{stability}, \emph{attractors} and
\emph{bifurcations}.

Typical examples for the stability and asymptotic stability of a
given set of solutions within a cosmology include the stability
problem of isotropic or homogeneous solutions with respect to
perturbations either in the given theory of gravity or in a larger
set of gravity theories (see, for example,  \cite{ba-ot83},
\cite{ba-so86}, \cite{co-fl93}).

The attractor properties of exact solutions of physical interest,
eg., inflationary, have occupied a great number of papers in the
literature. Many of these results describe analytic ways of how a
given solution approaches, or is approached by, another set of
solutions. However,  a rigorous general definition via dynamical
systems theory of the notion of cosmological attractor that will
prove useful in specific applications is still lacking. (The
recent book \cite{wa-el97} reviews some of these
problems in the modern language of dynamical systems and contains
basic results about equilibrium points, limit sets etc for the
(Bianchi/GR) family.)

Many cosmologies are formulated as a set of dynamical equations
with parameters for example, in the (Bianchi/GR) family there are
three such parameters describing the passage from one Bianchi type
to the next. Hence, the dynamics of cosmologies in many cases
present us with interesting \emph{bifurcation problems}. To our
knowledge bifurcation theory (cf. for instance, \cite{ha-ko91})
has not been considered in mathematical
cosmology up to now in any systematic way.

\section{Qualifying the approximations}
\begin{quote}
\emph{Most people, if you describe a train of events to them, will
tell you what the result would be. They can put those events
together in their minds, and argue from them that something will
come to pass. There are few people, however,  who, if you told
them a result, would be able to evolve from their own inner
consciousness what the steps were which led up to that result.
This power is what I mean when I talk of reasoning backward, or
analytically. (Sherlock Holmes)}
\end{quote}
 In the direction of
qualifying the approximations used a basic problem is to make
sense of the aforementioned BKL oscillatory pattern of approach
towards the initial cosmological singularity. We know that this is
a local and piecewise scheme for describing the general approach
to the singularity in, at least, the (BianchiIX/GR) category. It
is well-known that this oscillatory pattern of approach to the
singularity is disrupted by the inclusion of a scalar field
\cite{be-kh73,be99}, in which case a
monotonic approach to the singularity occurs, but the BKL
phenomenon returns in the additional presence of a vector field as
it was first noted in \cite{be-kh73}. A rigorous
analysis of this basic behaviour of (Bianchi/GR) cosmologies near
the spacetime singularity was first given by Bogoyavlenski and
Novikov in \cite{bo-no72} using the method of maximal conformal
compactification\footnote{This analysis, among many other important
results, is described in detail in the advanced book by
Bogoyavlenski, \cite{bo86}.}.

Aside from the notoriously difficult problem of deciding
\emph{how} local this behaviour is, we know that these same
patterns occur in many other cosmologies (taking into account
conformal dualities between different cosmologies which typically
transform them into the `kernel' category
($\cdot$/GR/scalarfields). This applies, for instance, to the
(BianchiIX /HDG)  category \cite{ba-co91} and certain
members of the (BianchiIX /ST) family. The existence of a BKL
scheme in higher dimensional (BianchiIX /GR) cosmologies is known
\cite{de-etal86} to be sensitive to the
number of spacetime dimensions. Recently, the BKL problem in the
context of certain (BianchiIX/String) cosmologies was considered
in \cite{ba-da97,da-he00}
and essentially similar behaviours appear. This
is due to the effect of scalar or vector fields on the vacuum
behaviour of these cosmologies as well as to the number of
dimensions.

However, the question of the genericity of the aforementioned
behaviour persists. Is the general solution of a generic vacuum
cosmology locally  a Mixmaster (oscillatory) one? Recent work
\cite{be-etal99}  shows that the answer to this question is
indeed `yes' in the (Inhomogeneous/GR) category.

A second, equally important, aspect of the issue of qualifying the
approximations relates to the observational cosmology program
developed by Ellis and his collaborators over the years
\cite{el-etal85,el-etal87}. Without the assumption that we do not occupy
a privileged position, the isotropy of the cosmic microwave
background does not imply an isotropic spacetime geometry. Thus
one proceeds by adding a new structure (a \texttt{CS}, a \texttt{TG}
etc), one at a time, and examines
what can be inferred about the geometry of the universe directly
from observations with minimal assumptions.

This program has been implemented within GR and it
would be very interesting to see what happens if a similar
approach is taken up in other cosmologies based on extensions of general relativity
such as, HDG or ST. For example, it can be shown \cite{el-etal85}
that without the use of a \texttt{TG} (in this case GR) one cannot
obtain a `distance-redshift' relation nor can prove, on the basis
of observations alone, that our spacetime is spherically symmetric
around us nor determine the sign of the curvature of the spatial
sections. It is unknown whether these results are valid in the
context of a HDG or a ST cosmology.

Hence a general `fitting' problem may be formulated for
cosmologies. Given a `lumpy' cosmology and another `ideal' one,
the question arises as to how to determine a best-fit between the
two. This question has been investigated in \cite{el-etal87}
within the context of general relativity. It is unclear whether
similar results are valid in other cosmologies.

\section{Ranges of validity}
\begin{quote}
\emph{The principal difficulty in your case lay in the fact of
there being too much evidence. What was vital was overlaid and
hidden by what was irrelevant. Of all the facts which were
presented to us we had to pick just those which we deemed to be
essential, and then piece them together in their order, so as to
reconstruct this very remarkable chain of events. (Sherlock
Holmes)}
\end{quote}
The last avenue of research, the range of validity of a given
cosmology, has at least two offshoots. The first can be well
described by a remark of D. Christodoulou \cite{cr99} (although made in
the asymptotically flat context, assuming that the strong
censorship conjecture turns out to be false) referring to the
question of whether or not a given system which develops
singularities at some finite time during its classical evolution
necessarily requires the exit from the classical phase and a
subsequent entrance to a quantum regime in order to have a
meaningful description of the evolution past the singularity:
\begin{quote}
\dots for, it is argued, that a physical system which initially
lies within the realm of validity of the theory would evolve into
a system which lies outside this realm and we would be compelled
to enter the domain of a quantum theory to obtain a valid
description. A breakdown of this type does indeed occur in the
Newtonian theory during stellar collapse to a neutron star or a
black hole. However, this is by no means the only possibility.
Another alternative is what very likely occurs in compressible
fluid flow past the starting point of shock formation. We then
have a new concept of solution for which complete regularity does
not hold but singularities are of a milder character, the shocks.
The subsequent evolution of these may be fully characterized by
the classical theory, even though their microscopic structure,
which resolves the discontinuities, is accessible only to a
molecular description\dots
\end{quote}
Perhaps the generic singularities in cosmology, as predicted by
the Hawking singularity theorems \cite{ha-el73},
are similar to shocks for many cosmologies. Do we then need a
theory of quantum gravity in order to describe the cosmological
evolution after or near cosmological singularities? What are the
true ranges of validity of our available cosmologies? Unravelling
the \emph{nature} of cosmological singularities is central to a
beginning of understanding possible answers to this question.

This brings up a new dimension to the range of validity issue one connected
with the view that there is no single cosmology which describes the universe
at all times but some cosmologies may be better adapted to some epochs than
others. This can be called the \emph{problem of cosmological cohesion}, that
is, to try to connect different cosmologies together to form a consistent
frame for cosmic history, \emph{a cohesive cosmology}, to compare with
observations and other constraints. For example, suppose that an
(FRW/GR/fluid) cosmology is valid after the Planck time onwards and that some
(Bianchi/String/Vacuum) cosmology holds well before that time. The
cosmological cohesion problem in this case is to connect the physically
meaningful classical solutions of the two cosmology branches into one cohesive
cosmology that would describe the entire cosmic history and be compatible with
observations and other constraints. Theorems that would describe precisely
what happens in such simple problems are lacking at present, but could greatly
contribute to our understanding of the behaviour of candidate cosmologies.

The second offshoot of the range of validity issue stems from the
fact that the solution spaces of different cosmologies are, in
general, not isomorphic. For example, consider two cosmologies
characterized by the same spacetime and matterfield structure but
different gravity theories\footnote{It follows from the conformal
equivalence theorem \cite{ba-co88} that the
solution space of, for instance, the family ($\cdot$/GR/scalarfield)
 is properly contained in the solution space of
the family ($\cdot$/HDG/scalarfield) as the latter contains the
($\cdot$/HDR/vacuum) family (obtained by setting the scalarfield
equal to zero) and this in turn is conformally equivalent to
($\cdot$/GR/scalarfield).}. How are their solution spaces related?
This question raises another one: How do we compare two
cosmologies? The importance of the possible answers to such
questions is obvious for we could transfer known results from one
cosmology to another augmenting in this way our knowledge of the
basic 'cosmology atlas'\footnote{The related issue of the
conservation laws of two comparable cosmologies is also of
interest as it is intimately connected to the problem of singling
out of two conformally related cosmologies one which contains the
true, physical metric that can be consistently used to measure
times and distances (see, \cite{fa-etal99} for a recent
review).}.

\section{Mathematical cosmology in retrospect}
\begin{quote}\emph{All science is cosmology, I believe. (Karl  Popper)}\end{quote}
Understanding the dynamics of general classes of cosmologies,
$\mathcal{C} =(\mathtt{CS},\mathtt{TG},\mathtt{MF})$, is the
central problem of mathematical cosmology. In its most general
form this problem is clearly intractable at present. There are
three equivalent ways of rephrasing this problem  splitting it
into three components as follows:
\begin{enumerate}
\item  The study of geometric and dynamical properties of
 a fixed class of spacetimes in the $(\mathtt{TG}%
,\mathtt{MF})$-space
\item  The study of a fixed theory of gravity in the $(\mathtt{CS}%
,\mathtt{MF})$-space
\item  The evolution properties of a fixed class of matterfields in the
$(\mathtt{CS},\mathtt{TG})$-space
\end{enumerate}
Each of the three aspects above represents a different projection
of the general problem of mathematical cosmology. As an exercise
the reader is invited to describe the differences  between the
three components of the general cosmological problem!

Further reductions and simplifications in each of these components
takes us too far afield and into other (most!) domains of pure or
applied mathematics. For example, consider `switching off' the
$\mathtt{TG}$ part in 3. Then one is left with an evolution
equation for a `matterfield' in a $\mathtt{CS}$ (and if we further
neglect the time coordinate we end up with a PDE for the
matterfield in some specified Riemannian space). Also pure
differential geometry can be thought of as the limit obtained from
1 when we switch off both $\mathtt{TG}$ and $\mathtt{MF}$ and
leave only the `space part' of the problem. Finally, any problem
in the calculus of variations can be arrived at from 2 by suitable
modifications. (Exercise!)

Theories of gravity, spacetimes and matterfields are the nuts and
bolts of the mathematical cosmologist in his/her attempts to
construct models that have the potential to consistently describe
the universe, the cosmologies. Mathematical cosmology is a vast
edifice in geometry, and dynamical systems theory plays a central
role in all attempts to unravel the behaviour of every possible
cosmology. Today  mathematical cosmologists have a well-defined
and consistent framework to exercise their imagination and special
mathematical skills in their efforts to replace current puzzling
issues awaiting for solution by new and more interesting ones. As
Poincar\'{e} once put it: 'mathematicians do not destroy the
obstacles with which their science is spiked, but simply push them
toward its boundary'.
\vspace{1.0cm}

\noindent I wish to thank Peter Leach and John Miritzis for useful discussions and
their critical readings of an initial version of this
manuscript.

\end{document}